\documentclass{ws-procs9x6}

\begin{document}

\title{A short introduction to \\
the fate of the $\alpha$-vacuum}

\author{HAEL COLLINS
\footnote{\uppercase{T}his work was done in collaboration with \uppercase{R}ich \uppercase{H}olman and \uppercase{M}atthew \uppercase{M}artin
and with support by the \uppercase{D}epartment of \uppercase{E}nergy  (\uppercase{DE}-\uppercase{FG}03-91-\uppercase{ER}40682).}}

\address{Department of Physics \\
Carnegie Mellon University \\ 
Pittsburgh, PA\ \  15213, USA\\ 
E-mail:  hcollins@andrew.cmu.edu}

\maketitle

\abstracts{
A free scalar field propagating in de Sitter space has a one parameter family of invariant states called the $\alpha$-vacua.  In an interacting theory, all except a unique state, the Bunch-Davies vacuum, produce non-renormalizable divergences.  This talk provides a brief introduction to the origin and the form of these divergences.
}

A striking difference between de Sitter and flat space-time is the richer vacuum structure of the former.  For a free, massive scalar field in Minkowski space, there exists a unique Poincar\' e invariant vacuum state.  Flat space also has a global time-like Killing vector, which allows us to define a Hamiltonian with respect to which the vacuum is the the lowest energy eigenstate.  In contrast, de Sitter space lacks any such global time-like Killing vector and so no globally conserved energy can be defined.  It is still possible to find states which are invariant under the connected part of the isometry group of de Sitter space, $SO(4,1)$, but there no longer is only a single such invariant state.  Instead, for a free scalar field there exists an infinite family of invariant states\cite{alphavac} called the $\alpha$-vacua, since they can be distinguished by a single complex number $\alpha$.  The $\alpha$-vacua have been used to explore in a wide set of problems associated with de Sitter space, such as the transplanckian problem of inflation\cite{transplanck} and the dS/CFT correspondence\cite{dSCFT}.

This talk describes a pathology that arises for an interacting scalar field theory in an $\alpha$-vacuum.\cite{fate,interalpha}  Its origin lies in the structure of the Green's function chosen to describe propagation in an $\alpha$-vacuum.  In a generic loop correction, the interference among different propagators produces terms in which no oscillatory suppression damps the high momentum region of the loop integration.  Loops which contain two or three propagators then lead to linear and logarithmic divergences respectively which cannot be removed by a counterterm of the form of those already present in the theory.  Only a unique choice of these vacua---known alternately as the Euclidean or Bunch-Davies vacuum\cite{bunch}---evades this class of divergences and yields completely renormalizable interacting theories in de Sitter space.

We begin with a free scalar field $\Phi(x)$ propagating in a classical de Sitter background coordinatized by 
\begin{equation}
ds^2 = {d\eta^2 - d\vec x^2\over\eta^2} ,
\qquad
\eta \in [-\infty,0] . 
\label{metric}
\end{equation}
We can define a vacuum state $|E\rangle$ to be the state annihilated by the operators $a_{\vec k}^E$ obtained by expanding the field in terms of its eigenmodes, 
\begin{equation}
\Phi(\eta,\vec x) = \int {d^3\vec k\over (2\pi)^3}\, \left[ 
U_k^E(\eta) e^{i\vec k\cdot \vec x}\, a^E_{\vec k} 
+ U_k^{E*}(\eta) e^{-i\vec k\cdot \vec x}\, a_{\vec k}^{E\dagger} \right] .
\label{eigenmodes}
\end{equation}
The mode functions $U_k^E(\eta)$ are the solutions to the second order Klein-Gordon equation.  One of the constants of integration is fixed by the requirement that $\Phi(x)$ and its conjugate momentum $\Pi(x) = \eta^{-2}\partial_\eta\Phi(x)$ satisfy the canonical commutation relation.  In flat space, the second constant of integration can be fixed unambiguously since the space-time has a global time-like Killing vector with respect to which we define positive and negative frequencies.  At short distances with respect to the natural curvature length of de Sitter space, the geometry also appears to be flat and we can thus use a prescription in which the mode functions are those which match onto the flat space modes with positive frequencies at short distances.  This prescription defines the Euclidean, or Bunch-Davies,\cite{bunch} vacuum, 
\begin{equation}
U_k^E(\eta) = {\sqrt{\pi}\over 2} \eta^{3/2} H_\nu^{(2)}(k\eta) ,
\qquad
\nu = \left[ {\textstyle{9\over 4}} - m^2 \right]^{1/2}
\label{Emodes}
\end{equation}
with $k \equiv |\vec k|$.  $m$ is the effective mass of $\Phi(x)$ in units where the Hubble constant is unity.  The Wightman function for this vacuum, $G_E(x,x') = \langle E | \Phi(x)\Phi(x')| E\rangle$, only depends on the de Sitter invariant distance between $x$ and $x'$.

In de Sitter space, the statement as to which are the positive and negative frequency modes cannot be made globally so the Euclidean vacuum is not the unique invariant state.  Mottola and Allen, as well as earlier authors, demonstrated\cite{alphavac} that the state $|\alpha\rangle$ annihilated by the Bogolubov transform, 
\begin{equation}
a_{\vec k}^\alpha = N_\alpha \bigl[ a_{\vec k}^E - e^{\alpha^*} a_{-\vec
k}^{E\dagger} \bigr] , 
\qquad
N_\alpha = \bigl[ 1 - e^{\alpha+\alpha^*} \bigr]^{-1/2} , 
\label{aalphadef}
\end{equation}
with Re$\, \alpha < 0$, also yields an invariant Wightman function, 
\begin{equation}
G_\alpha(x,x') = \langle \alpha | \Phi(x)\Phi(x')| \alpha\rangle 
= \int {d^3\vec k\over (2\pi)^3}\, 
e^{i\vec k\cdot (\vec x-\vec x')}\, G_k^\alpha(\eta,\eta') . 
\label{Awight}
\end{equation}
Equation (\ref{aalphadef}) induces a transformation of the mode functions, selecting a different linear combination of the solutions to the Klein-Gordon equation, 
\begin{equation}
U_k^\alpha(\eta) = N_\alpha {\sqrt{\pi}\over 2} \eta^{3/2} \bigl[ 
H_\nu^{(2)}(k\eta) + e^\alpha H_\nu^{(1)}(k\eta) \bigr] .
\label{Amodes}
\end{equation}
These are the mode functions associated with an expansion of $\Phi(x)$ with respect to the $\alpha$-vacuum, $|\alpha\rangle$.  Note that the Euclidean vacuum is itself an $\alpha$-vacuum, corresponding to the limiting case where $\alpha\to -\infty$.

The construction of the propagator in an $\alpha$-vacuum is more subtle.  The most straightforward definition of the Feynman propagator, 
\begin{equation}
G_F(x,x') \equiv i \langle\alpha |T(\Phi(x)\Phi(x'))| \alpha\rangle 
= \int {d^3\vec x\over (2\pi)^3}\, e^{i\vec k\cdot (\vec x-\vec x')} 
G^F_k(\eta,\eta') , 
\label{Feyndef}
\end{equation}
is to define it to be the Green's function associated with a single point source,
\begin{equation}
\left[ \nabla^2_x + m^2 \right] G_F(x,x') 
= {\delta^4(x-x')\over\sqrt{-g(x)}} . 
\label{KGFeyn}
\end{equation}
In terms of the Wightman functions, the propagator is given by 
\begin{equation}
G^F_k(\eta,\eta') 
= \Theta(\eta-\eta')\, G^>_k(\eta,\eta') 
+ \Theta(\eta'-\eta)\, G^<_k(\eta,\eta') , 
\label{Gkpmdef}
\end{equation}
with $G^>_k(\eta,\eta') = G^<_k(\eta',\eta) = i\, G_k^\alpha(\eta,\eta')$.  For a general value of $\alpha\not= -\infty$, this propagator produces divergences in the perturbative corrections of an interacting theory which cannot be canceled by a simple set of counterterms.\footnote{These divergences could indicate a fundamental pathology of the general $\alpha$-vacuum, but they may also signal that we should apply a different construction for the propagator.  In particular, adding a source at the antipode of $x$ in Eq.~(\ref{KGFeyn}) produces a theory without any non-renormalizable divergences.\cite{double}}

From the perspective of the Euclidean vacuum, the transformation defining the $\alpha$-vacuum in Eq.~(\ref{aalphadef}) introduces excitations to arbitrarily high momenta which, unlike a thermal distribution, are not damped as $k\to\infty$.  This modification of the theory for high momenta leads to a new set of divergences for an interacting theory in an $\alpha$-vacuum which appear in loop corrections.  At arbitrarily short distances the positive and negative frequency modes interfere so as to cancel the rapidly oscillating phases among some of the terms within the loop integral.  Without such a cancellation, these phases would have damped the contribution from the high momentum region through an appropriate $i\epsilon$ prescription.  This interference of phases is a specific feature of the propagator in the $\alpha$-vacuum and does not occur in the Euclidean case.  

\begin{figure}[ht]
\centerline{\epsfbox{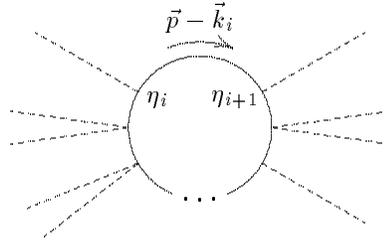}}
\caption{A general loop graph in an interacting scalar theory.
\label{loop}}
\end{figure}

To understand the origin of these divergences, consider a general one loop diagram as shown in Figure~\ref{loop}.  The loop contains $n$ vertices connected by $n$ propagators through which flow a common loop momentum $\vec p$.  The external legs are shown as dashed lines and contain the appropriate number of lines for the types of interactions included in the theory.  If the momentum leaving the loop at the $i^{\rm th}$ vertex is $\vec k_i-\vec k_{i-1}$, then the momentum flowing through the propagator connecting the $i^{\rm th}$ and the $(i+1)^{\rm th}$ vertices is $\vec p-\vec k_i$ as shown.  Expanding the Hankel functions in Eq.~(\ref{Amodes}) in the UV limit, the leading $p$-dependence of the Wightman functions associated with this propagator is 
\begin{eqnarray}
&&G_{p-k_i}^>(\eta_i,\eta_{i+1}) 
\label{wightUV} \\
&&\qquad\qquad\to
iN_\alpha^2 {\eta_i\eta_{i+1}\over 2|\vec p-\vec k_i|} 
\Bigl[ 
e^{-i|\vec p-\vec k_i|(\eta_i-\eta_{i+1})} 
+ e^{\alpha+\alpha^*} e^{i|\vec p-\vec k_i|(\eta_i-\eta_{i+1})} 
\nonumber \\
&&\qquad\qquad\qquad
- i e^\alpha e^{-i\pi\nu} e^{i|\vec p-\vec k_i|(\eta_i+\eta_{i+1})} 
+ i e^{\alpha^*} e^{i\pi\nu} e^{-i|\vec p-\vec k_i|(\eta_i+\eta_{i+1})} 
\Bigr] . 
\nonumber  
\end{eqnarray}
The prefactor determines the na\"\i ve UV behavior of the loop integral; since each propagator scales as $1/p$ for $p\gg k_i$, the integral of a loop containing $n$ propagators scales in the UV as 
\begin{equation}
\int^\Lambda d^3\vec p\, \prod_{i=1}^n G_{p-k_i}^\alpha(\eta_i,\eta_{i+1})
\sim 
\int^\Lambda {dp\over p^{n-2}} . 
\end{equation}
We only encounter a possible divergence if $n\le 3$.  Note that the $n=1$ case can be removed by a counterterm since in this case the loop is independent of any of the external momenta.

In the Euclidean case, the $\Theta$-functions in the propagators prevent the $p$-dependence of the phases from ever canceling in the UV in the product of all the propagators.\cite{sk}  For example, for $n=2$ the only products of loop Wightman functions which appear are $G_{p-k_1}^>(\eta_1,\eta_2) G_{p-k_2}^<(\eta_2,\eta_1)$ and its complex conjugate which oscillate rapidly as $p\to\infty$.  For the Euclidean vacuum, it is always possible to choose an $i\epsilon$ prescription so that the integral over three-momentum is finite for $\eta_1\not=\eta_2$.

For a generic $\alpha$-vacuum, the $\Theta$-function structure is identical; but since the phases depending on $p\eta_i$ and $p\eta_{i+1}$ appear with all possible relative signs in Eq.~(\ref{wightUV}), there will always exist terms in the product of $n$ loop propagators in which the $p$-dependent part of the phase completely cancels.  In the $n=2$ example, the product
\begin{eqnarray}
\quad&&\!\!\!\!\!\!\!\!\!\!\!\!\!\!
G_{p-k_1}^>(\eta_1,\eta_2) G_{p-k_2}^<(\eta_2,\eta_1) 
\label{ntwoalpha} \\
&&\to
e^{\alpha+\alpha^*} N_\alpha^4 
{(\eta_1\eta_2)^2\over 4|\vec p-\vec k_1| |\vec p-\vec k_2|} 
\nonumber \\
&&\quad\times\bigl[
e^{-i[ |\vec p-\vec k_1| - |\vec p-\vec k_2| ] (\eta_1-\eta_2)} 
+ e^{i[ |\vec p-\vec k_1| - |\vec p-\vec k_2| ] (\eta_1+\eta_2)} 
+ \hbox{c.c.} + \cdots
\bigr]
\nonumber  
\end{eqnarray}
contains terms whose $p$-dependent phases cancel for high loop momenta, $p\gg k_1,k_2$.  The loop integrand for these terms has no oscillatory suppression so that, when combined with the $p^2\, dp$ from the measure, the loop integral is linearly divergent.  The explicit $e^{\alpha+\alpha^*}$ prefactor shows why these divergences did not arise in the Euclidean ($\alpha\to -\infty$) vacuum.  Similarly, in the $n=3$ case we encounter a logarithmic divergence.

The dependence of the divergent part of the loop on the external momenta $\{k_1,\ldots k_{n}\}$ flowing into the loop is not the same as that of a counterterm that would generate a graph with the same external structure as Figure~\ref{loop} but with the loop shrunk to a point.\cite{fate}  It is in this sense that interacting theories in the generic $\alpha$-vacuum are non-renormalizable.

This result is shown in full detail for the example of the number density of Euclidean particles in the $\alpha$-vacuum for a theory with a $\Phi^3$ interaction in [4].  There we show that the expectation value of this number density produces a non-renormalizable linear divergence in the one loop correction.

The source of the divergences for an interacting theory in an $\alpha$-vacuum lies in the asymmetric treatment of the Wightman function and the Feynman propagator.  The Bogolubov transformation that produces the $\alpha$-vacuum can be regarded as a globally non-local transformation that defines the $\alpha$ mode functions to be a linear superposition of Euclidean mode functions evaluated at a point and at its antipode.  This non-locality is inherited by the Wightman function.  The time-ordering used to construct the Feynman propagator, indicated by the $\Theta$-functions in Eq.~(\ref{Gkpmdef}), does not contain any of this non-locality and is identical in the Euclidean and the general $\alpha$ cases.  This inconsistency suggests that the time-ordering should also be modified and that a renormalizable interacting field theory only results when the construction of the propagator consistently treats the global structure of de Sitter space.\cite{double}

\end{document}